\documentclass[prb,twocolumn,superscriptaddress,showpacs]{revtex4}
\usepackage{times}
\usepackage{amsmath}
\usepackage{amsthm}
\usepackage{amssymb}
\usepackage{amsbsy}
\usepackage{graphicx}
\usepackage{pstricks}
\usepackage{subfigure}

\begin{document}
\newcommand{\smallrech}{ \; \pspicture(0.2,0.1)
\psset{linewidth=0.025,linestyle=solid}
\psline[](0.1,0)(0.1,0.1)
\pspolygon[](0,0)(0.2,0)(0.2,0.1)(0,0.1)
\endpspicture\;}
\newcommand{\orients}{ \; \pspicture(0.01,0.01) \psarc{->}(-0.01,0.12){0.1}{10}{350} \endpspicture\;}

\newcommand{\smallrecv}{ \; \pspicture(0.1,0.2)
\psset{linewidth=0.025,linestyle=solid}
\psline[](0,0.1)(0.1,0.1)
\pspolygon[](0,0)(0,0.2)(0.1,0.2)(0.1,0.0)
\endpspicture\;}

\newcommand{\emptya}{  \; \pspicture(0.6,0.2)
\psdots[linecolor=black,dotsize=.10]  (0.15,0)(0.15,0.3)  (0.6,0.15)
\psset{linewidth=0.03,linestyle=dotted}
\psline[](0.3,0)(0.3,0.3)
\pspolygon[](0,0)(0.6,0)(0.6,0.3)(0,0.3)
\psset{linewidth=0.05,linestyle=solid}
\psline[linecolor=black](0,0)(0.3,0)
\psline[linecolor=black](0,0.3)(0.3,0.3)
\psline[linecolor=black](0.6,0)(0.6,0.3)
\endpspicture \;}

\newcommand{\emptyb}{ \; \pspicture(0.6,0.2)
\psdots[linecolor=black,dotsize=.10]  (0.45,0)(0.45,0.3)(0,0.15)
\psset{linewidth=0.03,linestyle=dotted}
\psline[](0.3,0)(0.3,0.3)
\pspolygon[](0,0)(0.6,0)(0.6,0.3)(0,0.3)
\psset{linewidth=0.05,linestyle=solid}
\psline[linecolor=black](0.3,0)(0.6,0)
\psline[linecolor=black](0.3,0.3)(0.6,0.3)
\psline[linecolor=black](0,0)(0,0.3)
\endpspicture \;}

\newcommand{\fulla}{ \; \pspicture(0.6,0.2)
\psdots[linecolor=black,dotsize=.10]   (0.15,0)(0.15,0.3)  (0.6,0.15)(0.3,0.15)
\psset{linewidth=0.03,linestyle=dotted}
\psline[](0.3,0)(0.3,0.3)
\pspolygon[](0,0)(0.6,0)(0.6,0.3)(0,0.3)
\psset{linewidth=0.05,linestyle=solid}
\psline[linecolor=black](0,0)(0.3,0)(0.3,0.3)(0,0.3)
\psline[linecolor=black](0.6,0)(0.6,0.3)
\endpspicture\;}

\newcommand{\fullb}{ \; \pspicture(0.6,0.2)
\psdots[linecolor=black,dotsize=.10]  (0.45,0)(0.45,0.3)  (0,0.15)(0.3,0.15)
\psset{linewidth=0.03,linestyle=dotted}
\psline[](0.3,0)(0.3,0.3)
\pspolygon[](0,0)(0.6,0)(0.6,0.3)(0,0.3)
\psset{linewidth=0.05,linestyle=solid}
\psline[linecolor=black](0.6,0)(0.3,0)(0.3,0.3)(0.6,0.3)
\psline[linecolor=black](0,0)(0,0.3)
\endpspicture \;}

\bibliographystyle{apsrev}

\title{Fermionic  quantum dimer and fully-packed loop models on the square lattice}

\author{Frank Pollmann}
\affiliation{Max-Planck-Institute for the Physics of Complex Systems, D-01187 Dresden, Germany}
\author{Joseph J. Betouras}
\affiliation{Department of Physics,
Loughborough University,
Loughborough, Leicestershire LE11 3TU, UK}
\author{Kirill Shtengel}
\affiliation{Department of Physics and Astronomy, University of California,
Riverside, CA 92521, USA}
\author{Peter Fulde}
\affiliation{Max-Planck-Institute for the Physics of Complex Systems, D-01187 Dresden, Germany}
\address{The Asia Pacific Center for Theoretical Physics, Pohang Korea}
\date{\today}

\begin{abstract}
We consider fermionic fully-packed loop and quantum dimer models which serve as effective low-energy models for strongly correlated fermions on a
checkerboard lattice at half and quarter filling, respectively. We identify a large number of fluctuationless states specific to each case, due to the fermionic statistics. We discuss the symmetries and conserved quantities of the system and show that for a class of fluctuating states in the half-filling case, the fermionic sign problem can be gauged away. This claim is supported by numerical evaluation of the low-lying states and can be understood by means of an algebraic construction. The elimination of the sign problem then allows us to analyze excitations at the Rokhsar-Kivelson point of the models using the relation to the height model and its excitations, within the single-mode approximation. We then discuss a mapping to a U(1) lattice gauge theory which relates the considered low-energy model to the compact quantum electrodynamics in 2+1 dimensions. Furthermore, we point out consequences and open questions in the light of these results.
\end{abstract}
\maketitle
\section{Introduction}
Strongly correlated electrons on lattices with geometric frustration can lead to new physical properties. These come essentially from the quantum fluctuations acting on an extensively degenerate collection of classical ground states. From that point of view, strongly interacting spinless fermions (or alternatively, fully spin polarized electrons) on a frustrated lattice present an interesting example of this kind of physics. A spectacular consequence for example is that at certain fillings, fractionally charged excitations have been predicted for the case of large nearest-neighbor repulsion on certain frustrated lattices.\cite{Fulde02} The subject is not purely academic since there is experimental evidence that electrons in systems which have the considered lattice structure can be strongly correlated.\cite{Kondo97, Walz2002} Furthermore, certain optical lattices with frustrated interactions can be produced by interfering laser beams and loaded with fermionic atoms. By varying the lattice constant, the strength of the correlations of the particles can be modified.\cite{Joksch05}

In the present work we focus on the problem of strongly correlated electrons on a checkerboard lattice at two different filling fractions. This lattice can be seen as a two-dimensional projection of the pyrochlore lattice (3D tetrahedra are represented by 2D squares whose diagonals and sides are on an equal footing as far as couplings are concerned). We will consider the case of strong nearest neighbor repulsion that favors the least possible number of fermions in each such square.
At half-filling, this number is two, and an effective low-energy model maps  onto a quantum fully-packed loop (FPL) model on a square lattice. Some parts of this work have been briefly discussed before.\cite{Pollmann06a,Pollmann06c} In the case of quarter-filling, a lowest potential energy arrangement corresponds to one fermion per square which can be mapped onto a hardcore dimer model. Here we supplement the study with further results and different aspects of the problem regarding the symmetries and conserved quantities, the sign problem and the corresponding gauge theories. We will discuss a number of consequences, especially related to the fermionic nature of the loops/dimers in this particular problem.
Quantum dimer models (QDMs) have have been studied quite extensively in the past two decades. \cite{Moessner08a}
Originally introduced in the context of quantum magnetism \cite{Rokhsar88}, they
became a focus of renewed interest following the discovery of a
gapped quantum liquid phase on a triangular lattice.\cite{Moessner01a}
It has been further established that a gapped topological phase
with deconfined excitations generically exists in 2D for QDMs on \emph{non-bipartite} lattices. \cite{Moessner01a,Nayak01a,Fendley02,Misguich02} Such a phase has been found to have an effective description in terms of a Z$_2$ gauge theory.\cite{Fradkin90,Moessner01b}
By contrast, on bipartite lattices (e.g.
a square lattice) there is a single deconfined quantum critical point separating phases of broken translation/rotation symmetry. \cite{Rokhsar88,Sachdev89,Moessner03b,Fradkin04}
Such a point is characterized by gapless excitation spectrum which in turn leads to logarithmic confinement at any finite temperature. In both cases, QDMs can be mapped onto loop models, with one significant difference: on a bipartite lattice the resulting loops can be naturally oriented while there is no natural notion of direction on a non-bipartite lattice. As a consequence,
an effective gauge theory for a bipartite case is now a U(1) theory (the oriented loops can be though of as field lines in the absence of sources).\cite{Fradkin90,Moessner01b} In addition, oriented loops naturally lend themselves to a height representation whereby they serve as contour lines. Several related models, such as the quantum six/eight-vertex models in 2D
have been shown to conform to the same dichotomy: a model with orientable loops is
either ordered or critical and is described by a U(1) gauge
theory while non-orientable quantum loops can result in a deconfined topological phase. \cite{Ardonne04,Chakravarty02,Shannon04,Syljuasen06a}
In all these models, the Hilbert space typically separates into disconnected subspaces corresponding to various conserved quantum numbers. When these quantum numbers are of non-local topological nature, such as winding numbers, the corresponding subspaces are referred to as topological sectors. While a  Z$_2$ topological phase is characterized by four such sectors and a corresponding four-fold ground state degeneracy on a torus, the U(1) models have a much larger number of conserved quantities.

All aforementioned models share one important feature: all off-diagonal matrix elements
connecting various quantum dimer/quantum loop states are non-positive (or can be made non-positive by a trivial gauge transformation such as associating a phase to dimers at certain locations).
  Within each disconnected sector the Hamiltonian dynamics is ergodic in the following sense: every two quantum states form a non-zero matrix element with \emph{some} power of the Hamiltonian. (Here we are not concerned with the issue of how such a power may depend on the system size -- a question essential for understanding the spectral gap or a lack thereof.)
The Perron-Frobenius theorem can then be applied to $\exp(-\tau H)$ for some positive $\tau$ within each disconnected subspace; it is follows that the
ground state for each sector is unique and
nodeless.\cite{Moessner01a,Castelnovo05b}

Much less is known about models with non-Frobenius dynamics which results in
some positive matrix elements. A striking difference between Frobenius vs.
non-Frobenius QDMs on the kagome lattice was found in
Refs.~[\onlinecite{Misguich02,Misguich03}]: while the conventional QDM exhibits a gapped
Z$_2$ topological phase, its counterpart with different signs has an extensive
ground state degeneracy. A different class of non-Frobenius lattice models  has also been studied recently: fermionic models with supersymmetric Hamiltonians; in a number of cases such systems have been shown to have an extensive
GS degeneracy.\cite{Fendley05b,Eerten05,Huijse08a,Huijse08b}.
From these examples, it appears that the non-Frobenius dynamics are often the source of additional quantum frustration resulting in the extensive ground state entropy.

The goal of the present manuscript is to present a number of results (and some open questions) related to another non-Frobenius model that was originally introduced in Refs.~[\onlinecite{Pollmann06c,Pollmann06d}].
The paper is organized as follows: We begin by introducing the considered model and derive an effective low-energy Hamiltonian in the following section. In Section~\ref{cons},  we begin by analyzing the
conservation laws specific to both quarter and half-filled cases, and the
constraints they impose on the quantum dynamics.  We
discuss the nature of the sign problem and present two different approaches to
alleviating it. In Section~\ref{sec:RK} we extend the model by adding an
extra term in the effective Hamiltonian that punishes flippable
configurations. In such an extended model, we analyze in
detail the Rokshar--Kivelson (RK) point and discuss the phase
diagram. In Section~\ref{sec:gauge}, we discuss the
gauge theory for both the quarter-filled and the half-filled cases. We conclude and summarize our results in Section~\ref{sec:conclusion}.

\section{The Model}
\label{sec:model}
\begin{figure}[thb]
  \begin{center}
\subfigure[ Two-particle exchange at half-filling]{\label{fig:FPL-2}
\includegraphics[width=0.45\columnwidth]{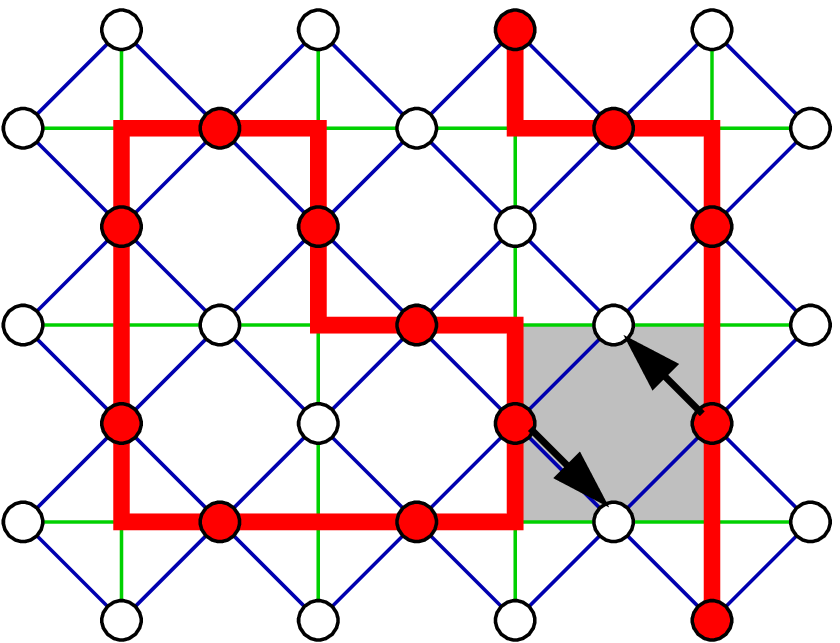}}
$\;\;$
\subfigure[ Three-particle exchange at half-filling]{\label{fig:FPL-3}
\includegraphics[width=0.45\columnwidth]{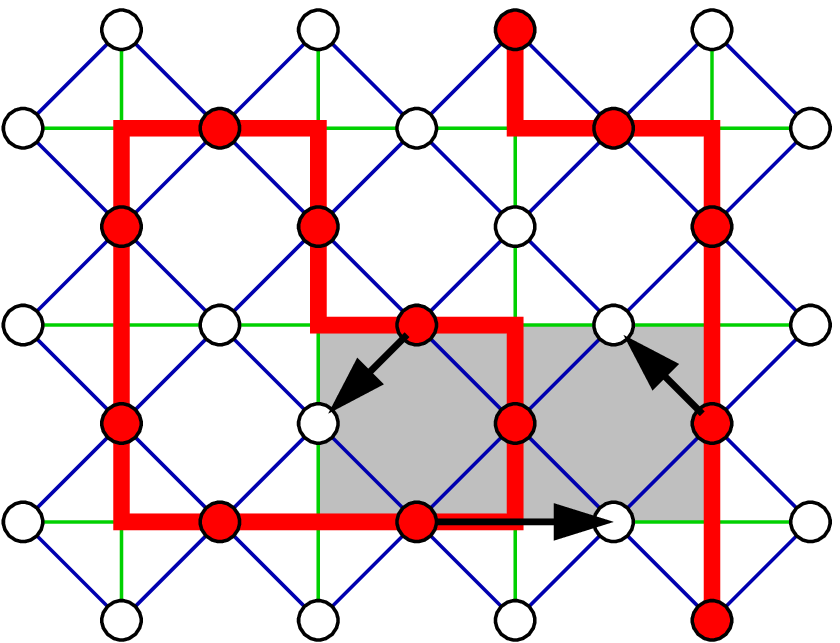}}
\\
\vskip 3mm
\subfigure[ Two-particle exchange at quarter-filling]{\label{fig:QDM-2}
\includegraphics[width=0.45\columnwidth]{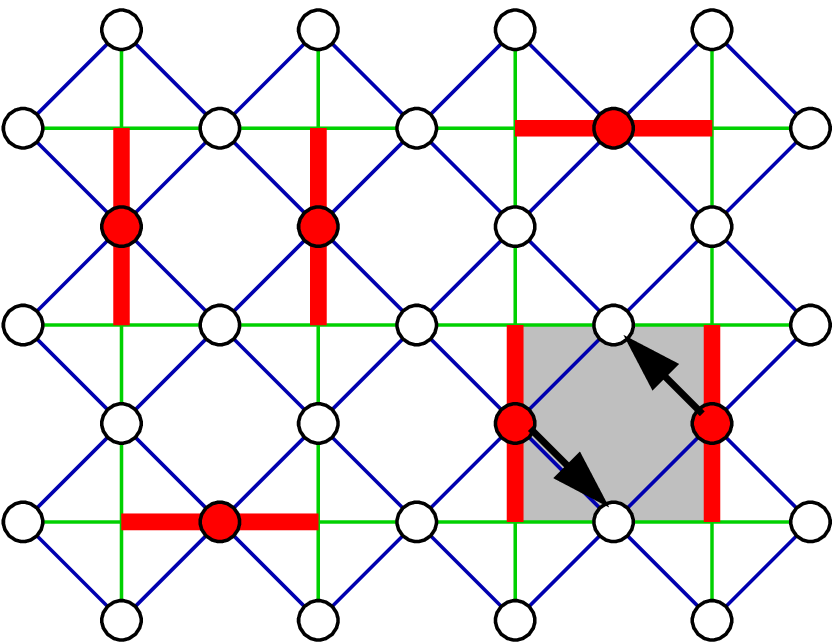}}
$\;\;$
\subfigure[ Three-particle exchange at quarter-filling]{\label{fig:QDM-3}
\includegraphics[width=0.45\columnwidth]{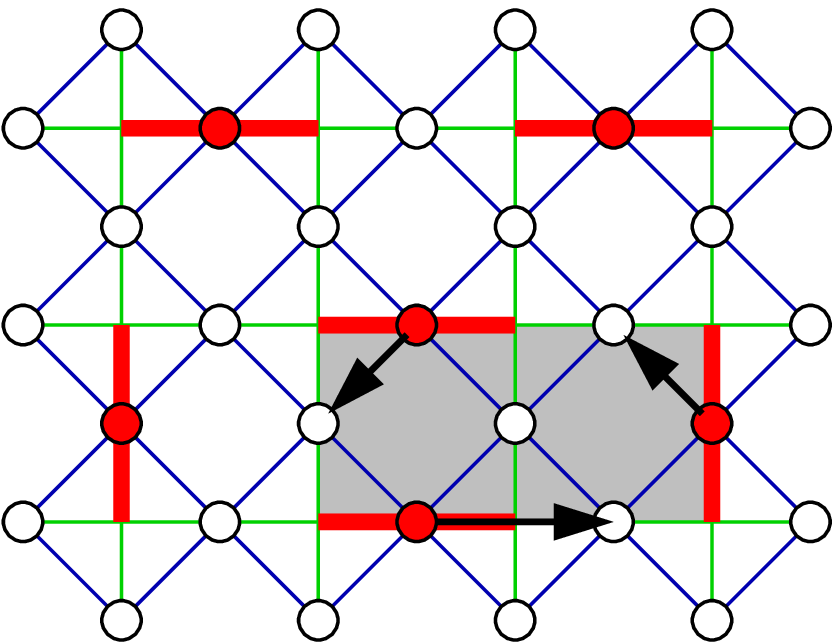}}
\end{center}
\caption{(a,b) --
A configuration of fermions satisfying the ``two per tetrahedron''
rule at half-filling and its representation in terms of FPLs.
(c,d) --
Configurations of fermions satisfying the ``one per tetrahedron''
rule at quarter-filling and their representations in terms of quantum dimers.
In both cases, ``flippable plaquettes'' corresponding to two- and three-particle ring exchanges are shaded (see text for
details).}
\label{fig:lattice-dimers}
\end{figure}

Our starting Hamiltonian for spinless fermions is
of the following form:
\begin{equation}
\label{eq:Hubbard1}
H=-t\sum_{\langle i\
j\rangle}\left(c_{i}^{\dag}c^{\vphantom{\dag}}_{j}+\text{H.c.}\right)+V\sum_{\langle i\
j\rangle}n_{i}n_{j},
\end{equation}
where the summation is performed over neighboring sites of a checkerboard lattice.  We are interested in the strongly interacting limit $V \gg t > 0$. Note that each site has six nearest neighbors as in the pyrochlore lattice.
The configurations favored by the interaction term depend on the doping, i.e., on the ratio of the number of particles to the number of the available sites. The fact that the particles are spinless fermions automatically prohibits double occupancy of sites. Furthermore, the interaction term favors uniform minimization of the number of such fermions on every crisscrossed square, a ``planar tetrahedron.''
We will be mostly interested in a case of half-filling whereby the potential energy term is minimized whenever there are two fermions on each
``planar tetrahedron'' as shown in Figure~\ref{fig:lattice-dimers}(a,b) (the tetrahedron rule
\cite{Anderson56}). The other case considered in parallel with the half-filling in this work is that of quarter-filling whereby the interaction energy is minimized by having one fermion per tetrahedron as shown in  Figure~\ref{fig:lattice-dimers}(c,d).\cite{Pollmann06a} In either case,
configurations satisfying these rules can be
represented by FPLs / dimers on a square lattice connecting the centers of the planar
tetrahedra. The particles are sitting in the middle of the bonds of the square lattice. At half-filling, the tetrahedron rule translates into a FPL covering. By the same token, at quarter-filling, the minimum interaction energy manifold is spanned by the dimer coverings of the square lattice such that each site belongs to one and only one dimer. In either case, all other configurations are separated by a potential energy gap of at least $V$.
Different FPL / dimer configurations are orthogonal to
each other since they correspond to different locations of fermions (here we assume that these locations simply label the orthonormal set of Wannier functions).

Both at half- and at quarter-filling, the kinetic term in the Hamiltonian (\ref{eq:Hubbard1}) moves particles between the neighboring sites thus
creating configurations in which the respective tetrahedron rules are violated. However, we are interested in the physics at the energy scale much smaller than $V$ and therefore  we should look at the correlated multi-particle hops (ring exchanges) instead. The lowest-order process that does not take the system out of the low-energy manifold is a two-particle ring exchange depicted in Figure~\ref{fig:lattice-dimers}(a,c). Such a process represents the dominant quantum dynamics for bosonic systems as well as for resonating valence bonds in spin systems.\cite{Rokhsar88} However, its amplitude vanishes identically for fermions: the relative signs of clockwise and counterclockwise processes are opposite as a consequence of the fermionic exchange statistics, and the two contributions cancel each other. Moreover, the same observation holds for all ring exchanges involving any \emph{even numbers} of fermions; the contributions from clock- and counterclockwise moves of an
even number of fermions similarly cancel. Therefore, the shortest allowed fermionic ring exchanges involve three particles rather than two (see Figure~\ref{fig:lattice-dimers}(b,d)). \footnote{Such cancelation of even ing exchanges is not a universal property but depends on the lattice: e.g., distorting a square lattice (where the dimers live) into a rhombic one will change the relative amplitudes of clockwise and counterclockwise exchanges and so that their contributions will not cancel any longer. } Note that if the spin of fermions is included, processes of order $t^2/V$ do no longer vanish: two electrons with opposite spin can resonate around empty plaquettes of the checkerboard lattice and lower the energy.\cite{poilblanc2007a, poilblanc2007b, trousselet2008}
By projecting the full Hamiltonian of Eq.~(\ref{eq:Hubbard1}) onto the Hilbert subspace restricted by the tetrahedron rules, we can perturbatively obtain an effective ring-exchange Hamiltonian that acts \emph{within} such a subspace.\cite{Runge04} To lowest
non-vanishing order in $t/V$, such effective Hamiltonian becomes:
\begin{multline}	
H^{(1/4)}_{\text{eff}} = g \sum_{\{\smallrech,\smallrecv\}}
\Big( \big|\emptya \big\rangle \big\langle\emptyb \big|+\text{H.c.}\Big)
\\
\equiv g\sum_p \Big(\big| A \big\rangle\big\langle \overline{A} \big| + \big| \overline{A} \big\rangle\big\langle {A} \big|
\Big)
\label{eq:ringexone-quarter}
\end{multline}
at quarter-filling and
\begin{multline}	
H^{(1/2)}_{\text{eff}} = g \sum_{\{\smallrech,\smallrecv\}}
\Big( \big|\emptya \big\rangle \big\langle\emptyb \big|+\text{H.c.}\Big)
\\
- g \sum_{\{\smallrech,\smallrecv\}}
\Big( \big|\fulla \big\rangle \big\langle\fullb \big|+\text{H.c.}\Big)
\\
\equiv g\sum_p \Big(\big| A \big\rangle\big\langle \overline{A} \big| + \big| \overline{A} \big\rangle\big\langle {A} \big|
- \big| B \big\rangle\big\langle \overline{B} \big| - \big| \overline{B} \rangle\langle {B} \big|\Big)
\label{eq:ringexone}
\end{multline}
at half-filling.
Here the sums are performed over all polygons of perimeter six and $g = 12
t^3/V^2$. As we have already mentioned, the actual fermions are located at the centers of the dimers. The different relative signs of two terms at a half-filling
signs result from the number of permuted fermions. It  is either even or
odd which in turn depends on whether the middle site of the hexagon is occupied or empty.  By contract, all ring exchanges come with the same sign in the case
of the quarter-filling, as they all correspond
to an empty middle site.

We should also point out that while the sign of the coupling constant $g$ is determined by the sign of $t$, it can be effectively changed by the following local gauge transformation:
multiply a states by a factor of $(-1)$ for each vertical dimer in columns numbered $4k$ and $4k+1$ (where $k\in\mathbb{Z}$); do the same for horizontal dimers replacing columns by rows. A simple check shows that all ring exchange terms in Eqs.~(\ref{eq:ringexone-quarter}) and (\ref{eq:ringexone}) change their sign as a result. (It is worth noting that any diagonal, i.e. potential energy terms that one might consider adding to these Hamiltonians would remain intact under this gauge transformation.) Therefore, despite all appearances to the contrary, the quantum dynamics in Eq.~(\ref{eq:ringexone-quarter}) is \emph{not} non-Frobebious. Such a gauge transformation, however, does not fix the relative sign of the different ring exchange terms in Eqs.~(\ref{eq:ringexone}) -- this issue will be addressed in the following section.

\section{Quantum Dynamics}
In this section, we discuss several aspects of the effective low-energy Hamiltonians which were derived in the previous section.
\label{cons}
\subsection{Conserved quantities}
\begin{figure}[thb]
  \begin{center}
\subfigure[]{\label{fig:dimer-heights}
\includegraphics[width=0.40\columnwidth]{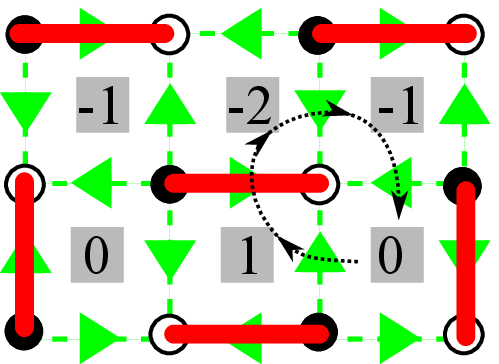}}
$\;\;\;\;\;\;\;\;$
\subfigure[]{\label{fig:6v-heights}
\includegraphics[width=0.40\columnwidth]{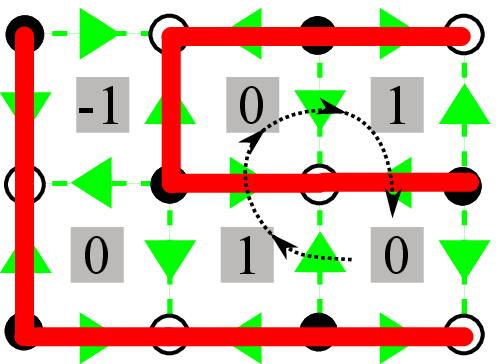}}
\end{center}
\caption{Recipe of how to map dimer and FPL coverings to a height model. First the links of the bipartite square lattice are oriented in such a way that the arrows are always pointing from the black to the white sublattice. Next we start from one arbitrary plaquette and set its value to zero. We then set the height value of the other plaquettes in the following ways. (a) Dimer model: If the traversed link is empty, we change the height by $\Delta h=1$ if the arrow is from left to right and by $\Delta h=-1$ if the arrow is from right to left. If the traversed link is occupied by a dimer, we change the height by $\Delta h=-3$ if the arrow is from left to right and by $\Delta h=3$ if the arrow is from right to left. The dashed line shows one possible path assigning the heights to the plaquettes (b) FPL model: The rules are nearly identical to those for dimers except that $|\Delta h|=1$ for both occupied and empty links. Since an FPL state is always an overlap of two dimer states, one can also obtain the same heights by simply adding those corresponding to the dimer states and dividing the result by two.}
\label{fig:height}
\end{figure}

Our first observation is that both at quarter and half-filling, the low
energy Hilbert space of our model can be represented by height
configurations. This follows from the aforementioned geometric
representations of the ground states of the potential energy term in
Eq.~(\ref{eq:Hubbard1}): in the case of quarter-filling they are the dimer
coverings of the square lattice while for the case of half-filling they form
FPL on the same lattice, which are isomorphic to
six-vertex configurations.\cite{Lieb72},\footnote{The
mapping between FPL and six-vertex configurations is done by
subdividing the square lattice into even
and odd sublattices and orienting occupied bonds from odd to even
sites, with the opposite orientation for vacant bonds.} Both, dimer coverings
and six-vertex configurations can in turn be mapped onto configurations of an interface
-- a so-called height representation, also known as a solid-on-solid (SOS)
 model.\cite{van_Beijeren77,Blote82,Levitov90} Figure~\ref{fig:height}(a,b) show
one of the ways of assigning heights to the plaquettes of the square lattice
for both dimer coverings and FPL configurations.

The quantum dynamics that preserves these geometric constraints then results in
quantum fluctuations of the local height variable. Hence our model is a variant
of a quantum height model.\cite{Henley97,Henley04}
Let us begin by reiterating some results known from studies of bosonic quantum dimer models\cite{Henley97,Henley04} and quantum six-vertex models.\cite{Chakravarty02,Ardonne04,Syljuasen06a} A conservation law which is valid for the ring exchanges of any finite length is the conservation of
the global slope of the height field. The global slope variable  has two components,
$\kappa_x(y)$ and $\kappa_y(x)$ in $x$ and $y$ directions, which
form a set of conserved numbers. (In case of periodic boundary conditions, these can be immediately translated into
the \emph{winding numbers}.)
There are two immediate consequences of this conservation law. The first
one concerns the ergodicity of the quantum dynamics: even for arbitrary local
ring exchanges, the configurations corresponding to the different topological
sectors (i.e., different overall slopes) cannot be mixed.  Secondly, certain
configurations, namely those corresponding to a maximal possible slope in
at least one direction must be frozen since any local change in the height field
would result in exceeding the maximal possible value of the step somewhere
around the affected plaquette(s).

For the fermionic variants of the FPL and dimer models, there are two important differences which result from fermionic statistics. Firstly, only ring exchanges involving \emph{odd numbers} of particles are allowed (as discussed in Section \ref{sec:model}). In particular, the smallest
resonating plaquette has perimeter six, rather than four, as captured by Eqs.~(\ref{eq:ringexone}) and (\ref{eq:ringexone-quarter}), and thus the height updates involve \emph{two} neighboring plaquettes rather than a single one. From the point of view of the fundamental conserved quantity of quantum height models -- the global slope $\kappa_x(y)$ and $\kappa_y(x)$ of the height field -- this difference is immaterial since the overall slope cannot be changed by any local dynamics.
In addition there is an accidental symmetry in our model, specific to the
quantum dynamics in our case: it separately conserves the total numbers of
vertical and horizontal dimers (moreover, it conserves the number of
horizontal dimers in even and odd rows separately, and similarly for the
vertical dimers). Hence, all states can be classified with respect to these
quantum numbers which are \emph{different} from the aforementioned conserved
slopes. Notice that these additional conservation laws will not hold if
higher-order ring-exchange terms are added to the effective Hamiltonian.

Secondly, the sign of the ring-hopping term in the fermionic FPL model depends on the
occupancy of the middle bond. Both differences have important
implications: the first one -- for the ergodicity of the quantum dynamics,
while the second one leads to a sign problem in the quantum Monte Carlo dynamics. As we will see in the following subsection, the sign problem can be cured in certain cases. We note that the overall sign of $H_{\text{eff}}$ (i.e.,
the sign of $g$) in Eqs.~(\ref{eq:ringexone}) and (\ref{eq:ringexone-quarter}) is a matter of convention (as
has been first pointed out in Ref.~\onlinecite{Rokhsar88} for the related bosonic QDM).
Introducing a factor of $i$ for each vertical dimer in an odd row of the square lattice
and for each horizontal dimer in an odd column changes this sign.

\subsection{Sign problem: Topological approach}
\label{sign}
\begin{figure}[bht]
\includegraphics[width=80mm]{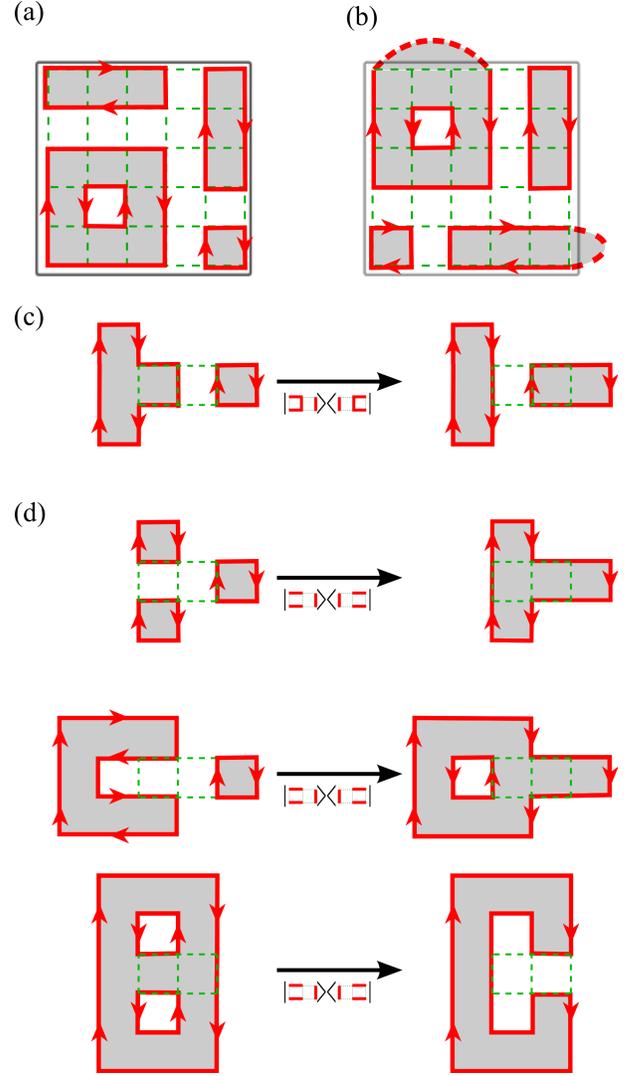}
\caption{(a, b) Representation of configuration as fully-packed directed loops. (c) Action of a ``non sign-changing'' process of the  effective Hamiltonian  (d) Three different actions of ``sign changing'' processes on the topology of loop configurations.}
\label{fig:loops}
\end{figure}
We will now address the sign problem manifested in the opposite signs of the $A\leftrightarrow
\overline{A}$ and $B\leftrightarrow \overline{B}$ terms in
Eqs.~(\ref{eq:ringexone}) and show that it can be avoided in certain (but not all) cases. Let us consider the surface with an
even number of fermions remaining fixed at the boundary (``fixed'' boundary
conditions). Then fermion configurations are represented by closed loops as
well as arcs terminating at the boundary. With no fermions at the
 boundary, there are closed loops only, as shown in Figure~\ref{fig:loops}(a). We add orientations to the loops  as follows: (i) color the areas separated by the loops white
and grey, with white being the outmost color; (ii) orient all loops so that
the white regions are always to the right. In the presence of arcs, Figure~\ref{fig:loops}(b), we can choose how to close them on the
outside without intersections (the outside region has no dynamics). Letting
white be the color at infinity, we orient the loops as described above.

We now consider how the effective Hamiltonian affects the loops covering. We first notice that the $B\leftrightarrow \overline{B}$ processes do not change the loop
topology (or their number) while $A\leftrightarrow \overline{A}$ flips always do, with the possibilities presented schematically in
Figure~\ref{fig:loops}(c,d). The relative signs resulting from the $A\leftrightarrow
\overline{A}$ flips are consistent with multiplying each loop
configuration by $i^l (-i)^r$, where $r$
($l$) is the number of the (counter-) clockwise winding loop.
 Hence, by simultaneously changing the sign of the $A\leftrightarrow
\overline{A}$ flips and transforming the loop states $|\mathcal{L} \rangle \to
i^{l(\mathcal{L})} (-i)^{r(\mathcal{L})} |\mathcal{L} \rangle$, we cure the
sign problem thus making the system effectively bosonic.

\begin{figure}[thb]
(a)\includegraphics[width=80mm]{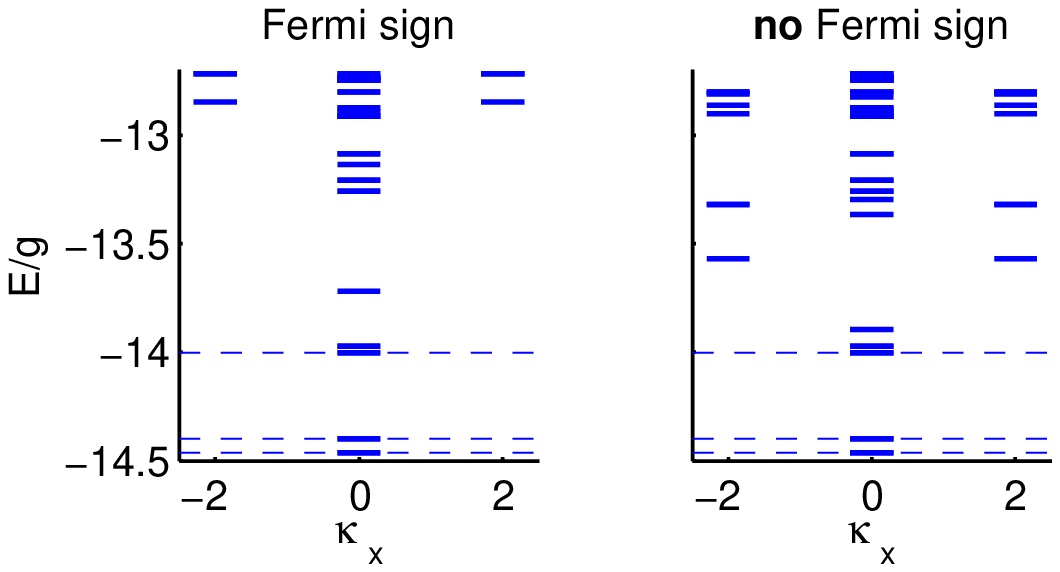}\\
~\\
~\\
(b)\includegraphics[width=80mm]{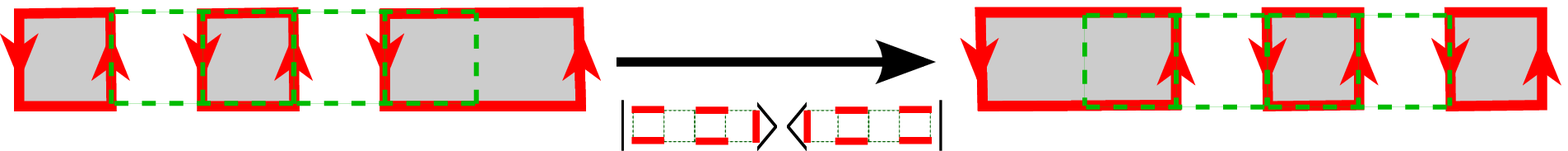}
\caption{(a) GS energy and energies of lowest excited states of the
effective Hamiltonian in sectors with different slopes ($\kappa_x$, $0$). Left
panel: Data from an exact diagonalization of $H_{\text{eff}}$ on a 72 site
cluster where the Fermi sign is taken into account. Right panel: Same data
from a calculation with excluded Fermi signs. (b) A counter example of a fifth order process which changes the sign of the wave function but not the topology of the loops covering.}
\label{fig:parabola}
\end{figure}

Note that this construction need not work for the periodic boundary conditions. Firstly, only
even-winding sectors on a torus allow two-coloring. Secondly, even for
even windings,  we can start from a given configuration and find sequences of hexagons-flipping processes which invert the coloring when returning to the initial configuration (not shown here). This means that it is possible to dynamically reverse the coloring while and thus the sign gauging fails for these cases. It appears though that for the periodic boundary conditions on \emph{even} tori (preserving
the bipartiteness of the lattice), the lowest-energy states belong to the
sector where such transformation works, as confirmed by the exact
diagonalization results (Figure~\ref{fig:parabola}(a)).

The presented non-local loop-orienting construction is furthermore restricted to the
ring-exchange processes of length six. The next higher non-vanishing terms result from processes involving five fermions hopping around a polygon of length 10. The fact that the above derived rules for the sign change due to the processes involving three fermions are violated can be seen from a counter example shown in Figure~\ref{fig:parabola}(b). Here the five fermions hopp around an odd number of fermions in the middle and thus  yield a sign change. However, the fifth order process here does \emph{not} change the number or the orientation of the loop. Consequently, the sign gauging based on the topology of the loops breaks down once these terms are included.

\subsection{Sign problem: Algebraic approach}

So far, we have attacked the sign problem using topological and computational arguments.
In order to complement these approaches with an algebraic expression of the gauge transformation,  we are
seeking a possible way of representing the Hamiltonian (\ref{eq:ringexone}) with
the kinetic part in such  way that the ``sign problem'', which is manifested
in the relative sign of the two terms, disappears.

\begin{figure}[hbt]
\includegraphics[width=50mm]{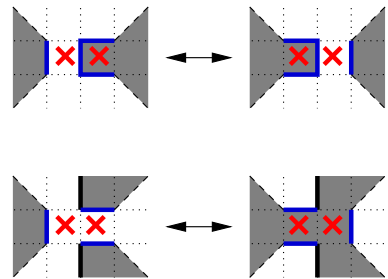}
\caption{Two distinct ``plaquette flips'' and corresponding colorings of
the plane. The two moves are different by the relative ``$-$'' sign. The
dashed diagonal lines represent either of the two equivalent (for the purpose
of our argument) locations of a dimer along one of the adjacent sides of a
square. The red crosses indicate the the double plaquette around which the loop segments resonate.}
\label{fig:dynamics}
\end{figure}

We use grey-and-white coloring of the plane as has been discussed above.
The two distinct possibilities of coloring a flippable rectangle (up to an overall color reversal) are shown
in Figure~\ref{fig:dynamics}. Notice that in both cases, both 
plaquettes forming a flippable rectangle (marked by the crosses) change their colors as a result of a ``flip''. The key observation though is that in one case
the color of these plaquettes is different and in the other it is the
same.

Let us define an SU(2) ``isospin'' $\sigma$ defined on each plaquette so that
the eigenvalue $\sigma^z=\pm 1$ indicates whether the plaquette is white or
grey. We can then rewrite the Hamiltonian (\ref{eq:ringexone}) in terms of
these new variables:
\begin{equation}
H_{\text{kin}} =
g \sum_{\langle i,j\rangle} \sigma_i^z \sigma_j^z \sigma_i^x \sigma_j^x
\Delta_{\text{flip}}(\sigma^z).
\label{eq:ringexsigma1}
\end{equation}
We will now discuss the terms and show that (\ref{eq:ringexsigma1}) is a good representation of the original effective Hamiltonian. The operators $\Delta_{\text{flip}}(\sigma^z)$ is a
projector which annihilates a state unless a legal flippable
configuration exists in the plaquette pair $(i,j)$, in which case its
eigenvalue is 1. The explicit form of $\Delta_{\text{flip}}(\sigma^z)$ is
complicated -- but what is important, it can only depend on the
$\sigma^z$'s of the two affected plaquettes and their neighbors.
(This is just a formal consequence of the fact that if we can look at the
coloring and tell a flippable configuration from a non-flippable, hence there
should be an operator that can do the same thing by ``measuring'' such
coloring which in turn is uniquely determined by the values of $\sigma^z$). If $\Delta_{\text{flip}}(\sigma^z)$ returns 1 upon acting on a
configuration, then the operators $\sigma_i^x \sigma_j^x$ simply flips the two isospins
(i.e. flips the colors of both plaquettes), i.e., it inverts the colors of the two plaquettes. The operator $\sigma_i^z \sigma_j^z$
determines the overall sign of this flip by distinguishing the two cases shown
in Figure~\ref{fig:dynamics}. In other words, it yields the sign change which occurs in the $B$ processes and represent the sign problem that we are
attempting to tackle!

The resolution of the sign problem in the chosen representation is actually trivial:
$\sigma^z \sigma^x = i \sigma^y$ and therefore
\begin{equation}
H_{\text{kin}} =
- g \sum_{\langle i,j\rangle} \sigma_i^y \sigma_j^y
\Delta_{\text{flip}}(\sigma^z).
\label{eq:ringexsigma2}
\end{equation}

But the only ``physical'' direction in the isospin space is $z$ -- it
corresponds to the plaquette being grey or white. The choice of $x$ and $y$
directions is a ``gauge'' choice. Hence we can always do a rotation around
the $z$ axis, so that $y\to x$ (in fact, we should rotate $y\to \pm x$ for
the sites of the even/odd sublattice in order to consume the extra minus
sign in Eq.~(\ref{eq:ringexsigma2})). The result is
\begin{equation}
H_{\text{kin}} =
g \sum_{\langle i,j\rangle} \sigma_i^x \sigma_j^x
\Delta_{\text{flip}}(\sigma^z).
\label{eq:ringexsigma3}
\end{equation}
which is precisely Eq.~(\ref{eq:ringexsigma1}) but without the different signs!
Notice that this ``spin rotation'' affects the phases of the wave functions
(written in $\sigma^z$ representation). This construction coincides with the ``loop
transformation'' we have presented earlier.

The non-local nature of this transformation is obvious from the fact that
$\sigma$'s are non-local operators:
\begin{equation}
\sigma_i^z = \exp{ \left[i\pi \sum_{k\in \mathcal{C}}c_k^\dag c^{\vphantom{\dag}}_k \right]}
\label{eq:Wigner-Jordan}
\end{equation}
where $\mathcal{C}$ is a cut going from the given plaquette to the boundary
(assumed white for simplicity), $c_k^\dag$ and $c_k$ are creation and annihilation operators for fermions at wavenumber $k$.
$\sum_{k\in \mathcal{C}}c_k^\dag c_k^{\vphantom{\dag}}$
simply counts the number of dimers crossing this cut (i.e., the number of
fermions on the cut). This is actually identical to the Jordan--Wigner
transformation in 1D! This suggests that our model is ``secretly'' 1D-like,
which explains why the fermionic sign can be gauged away. Notice that this nice property appears to be a consequence of restrictive three-particle dynamics described by Eq.~(\ref{eq:ringexone}); adding five-particle ring exchanges seems to lead to a genuine sign problem that cannot be similarly fixed by a gauge transformation.

\section{Rokhsar-Kivelson (RK) Point}
\label{sec:RK}

Numerous examples demonstrate that a Hamiltonian containing only kinetic, plaquette flip (or ring exchange) terms leads to an ordered phase stabilized by the ``order by disorder'' mechanism. On a square lattice, both the usual QDM and the quantum 6-vertex (FPL) models are at their plaquette phase\cite{Leung96,Syljuasen06b,Shannon04,Syljuasen06a} characterized by a checkerboard pattern (half as dense for the QDM case) of resonating plaquettes which break translational but not rotational symmetry of the lattice. A vary similar plaquette phase occurs on a honeycomb lattice\cite{Moessner01c}, where the dimer model and FPL models are simply dual to one another. Moreover, a somewhat more complicated broken symmetry state of the quantum dimers, dubbed $\sqrt{12}\times\sqrt{12}$, has been found on a non-bipartite, triangular lattice.\cite{Moessner01b,Ralko05} In the fermionic case at quarter and half-filling the Hamiltonian (\ref{eq:ringexone})
has also been shown to have an ordered ground state and fractionally charged, ``weakly'' confined excitations.\cite{Pollmann06b, Pollmann06f}

In order to explore the possible quantum phases with fermionic ring exchanges, we follow the example of  Ref.~\onlinecite{Rokhsar88} and extend our model's parameter space by adding a potential term as follows:
\begin{multline}
 \label{eq:ringexqdm}
H_{\text{eff}}^{\prime (1/4)} = \sum_p \left[-g\left(\left| A \rangle
\langle \overline{A} \right| + \left| \overline{A} \rangle
\langle {A} \right|
\right)
\right.
\\
\left.
+ \mu \left( \left| A \rangle
\langle {A} \right| + \left| \overline{A} \rangle
\langle \overline{A} \right|\right)\right]
\end{multline}
for the quarter-filled case and
\begin{multline}
 \label{eq:ringextwo}
H_{\text{eff}}^{\prime (1/2)} = \sum_p \left[-g\left(\left| A \rangle
\langle \overline{A} \right| + \left| \overline{A} \rangle
\langle {A} \right|
+ \left| B \rangle
\langle \overline{B} \right| +\left| \overline{B} \rangle
\langle {B} \right|\right) \right.
\\
\left.
+ \mu \left( \left| A \rangle
\langle {A} \right| + \left| \overline{A} \rangle
\langle \overline{A} \right|
+ \left| {B} \rangle
\langle {B} \right| + \left| \overline{B} \rangle
\langle \overline{B} \right|\right)\right]
\end{multline}
for the half filled case.
Here we use the short-hand notations introduced in Eqs.~(\ref{eq:ringexone-quarter}) and (\ref{eq:ringexone}) and in both cases applied the gauge transformations discussed in Section~\ref{sign} to change the sign of coupling constant in front of the $A\leftrightarrow
\overline{A}$ terms. The Rokhsar-Kivelson (RK) point corresponds to $\mu=g$ where the Hamiltonian can be written as a sum of projectors
\begin{multline}
 \label{eq:RK}
H_{\text{eff}}^{\prime}|_{g=\mu} = g\sum_p \Big[\Big(| A \rangle
- | \overline{A} \rangle \Big)
\Big(\langle  A | - \langle  \overline{A} | \Big)
\Big.
\\
+ \Big.
\Big(| B \rangle - | \overline{B}
\rangle \Big)
\Big(
\langle  B | - \langle  \overline{B} | \Big)
\Big].
\end{multline}
for the half-filled case; only the the first line, i.e., $A$ terms are present for the quarter-filling. The important feature of this Hamiltonian is that it is positive semidefinite, hence (i) a zero-energy state is automatically a ground state and (ii) such a state must be
annihilated simultaneously by all projectors. The consequence of the fact that the gauge-transformed Hamiltonian is of the Frobenius type is that for each subspace of states connected by the quantum dynamics, the equal amplitude superposition of such states is a unique ``liquid'' ground state. There is, however, a separate class of additional ground states -- those not connected to any other states. These states by definition have no flippable configurations and hence are trivially annihilated by Hamiltonian~(\ref{eq:RK}).
We note that such ``frozen'' states remain zero-energy states of Hamiltonians~(\ref{eq:ringexqdm}),(\ref{eq:ringextwo}) for all values of $\mu$, and remain the ground states for all $\mu > g$ as these Hamiltonians retain their positive semidefinite nature in this region.

\begin{figure}[thb]
  \begin{center}
\subfigure[]{\label{fig:stag1-quarter}
\includegraphics[width=0.35\columnwidth]{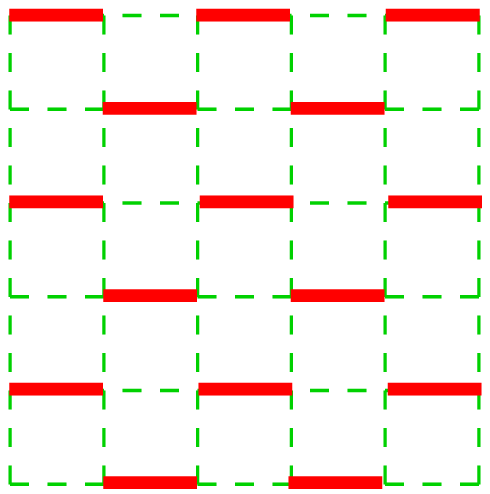}}
$\;\;\;\;\;\;$
\subfigure[]{\label{fig:stag1-half}
\includegraphics[width=0.35\columnwidth]{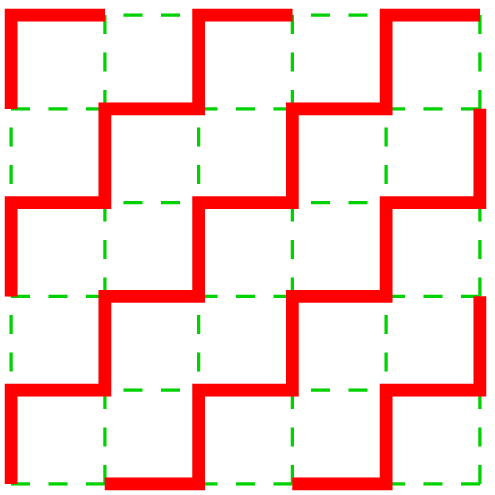}}
\\
\vskip 3mm
\subfigure[]{\label{fig:column-quarter}
\includegraphics[width=0.35\columnwidth]{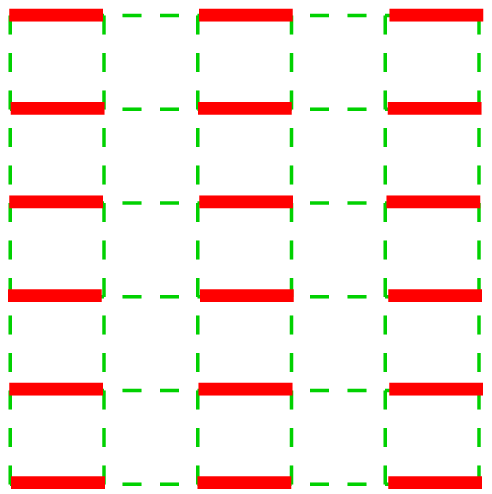}}
$\;\;\;\;\;\;$
\subfigure[]{\label{fig:column-half}
\includegraphics[width=0.35\columnwidth]{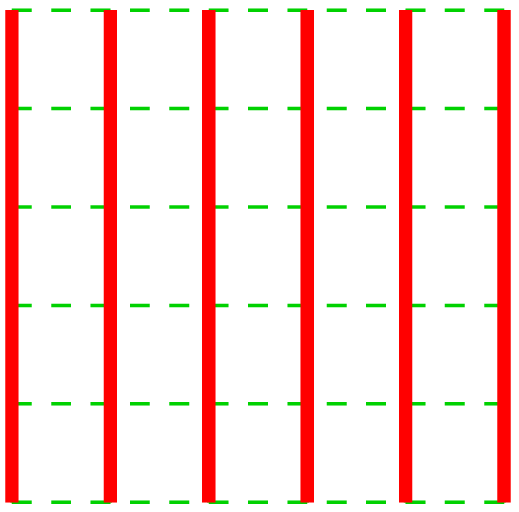}}
\\
\vskip 3mm
\subfigure[]{\label{fig:generic-quarter}
\includegraphics[width=0.35\columnwidth]{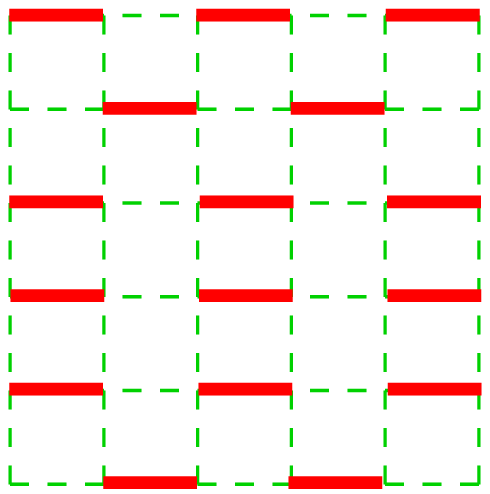}}
$\;\;\;\;\;\;$
\subfigure[]{\label{fig:generic-half}
\includegraphics[width=0.35\columnwidth]{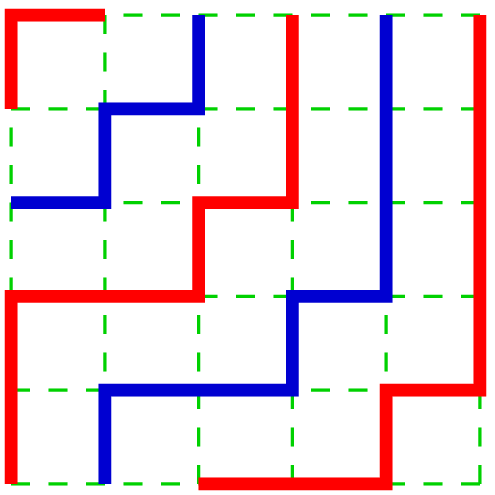}}
\\
\vskip 3mm
\subfigure[]{\label{fig:frozen-plaquettes}
\includegraphics[width=0.35\columnwidth]{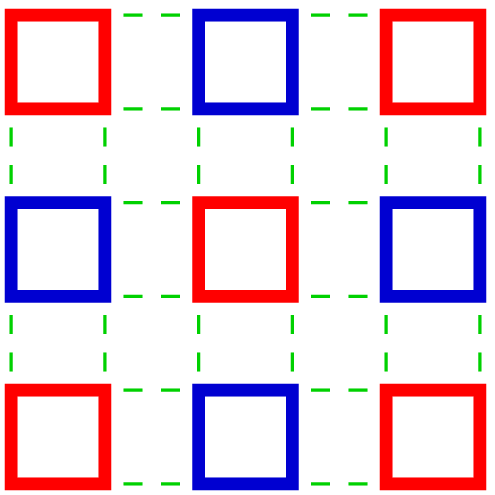}}
$\;\;\;\;\;\;$
\subfigure[]{\label{fig:frozen-rings}
\includegraphics[width=0.35\columnwidth]{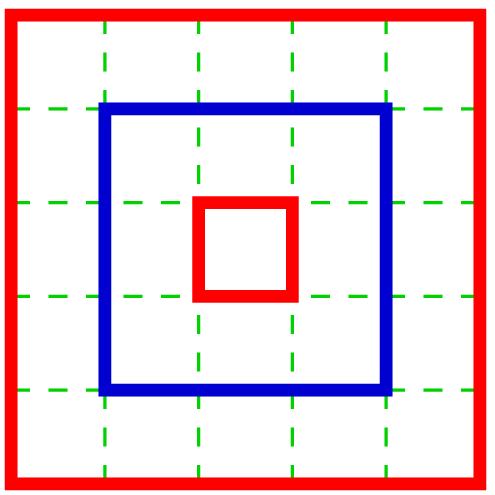}}
\end{center}
\caption{
Fragments of possible ``frozen'' states:
(a,b) are fluctuationless under \emph{any} local ring exchanges at quarter-filling (a) and half-filling (b);
(c,d) are maximally flippable for bosonic, but fluctuationless for fermionic ring exchanges;
(e,f) are the examples of generic fluctuationless configurations of fermions at both quarter-filling (e) and half-filling (f);
(g,h) are other examples of fluctuationless configurations of fermions at half-filling.
}
\label{fig:frozen}
\end{figure}

While the fermionic nature of our original Hamiltonian has been seemingly gauged away, leaving us with a Frobenius type, we should recall that it is still manifest in the fact that only odd ring exchanges are allowed. This has a profound consequence for frozen states.
All bosonic frozen states (i.e., frozen under a single plaquette flip dynamics) remain frozen in our case. This is immediately obvious on a superficial level: all rectangles flippable by either $A\leftrightarrow
\overline{A}$ or $B\leftrightarrow \overline{B}$ terms contain a flippable plaquette, hence a state with no flippable plaquettes is impervious the dynamics of Hamiltonians~(\ref{eq:ringexqdm}) or (\ref{eq:ringextwo}). On a deeper level, bosonic frozen states remain frozen under fermionic dynamics since
they are unaffected by \emph{any} local ring exchange (see Figures~\ref{fig:frozen}(a,b)). In the height language, these states maximize the slope in some direction, making any
local dynamics impossible. There are typically $\sim2^L$ such configurations, their exact number may depend on the boundary conditions. Such counting is easiest at half-filling where one may maximize the slope of an $L\times L$ system in in either $\pm x$ or $\pm y$ direction, which still leaves one with a freedom of
 $2^L$ possible arrangements of steps in the other direction
resulting in a $4\times 2^L$ number of frozen states.
In the quarter-filled (QDM) case there are four \emph{staggered} states of the type shown in Figure~\ref{fig:stag1-quarter} that maximize the slope in either $\pm x$ or $\pm y$ direction, with the slope in the other direction automatically fixed to zero. Additional frozen states can be obtained by creating diagonal domain walls between horizontal and vertical staggered states. Once again, all these states remain frozen under any local dynamics, fermionic or bosonic.

The non-trivial consequence of the fermionic dynamics is the appearance of a large number of additional frozen states that is not related to maximizing the slope, in fact many of these states are in the flat sector. Two such examples are presented in see Figures~\ref{fig:frozen}(c,d). Both the columnar state of quantum dimers (Figure~\ref{fig:column-quarter}) and the analogous state of FPL (Figure~\ref{fig:column-half}) actually maximize the number of flippable plaquettes, yet both are fluctuationless under \emph{any} fermionic dynamics. These states (along with staggered states) are members of a larger family of frozen fermionic states shown in Figures~\ref{fig:frozen}(e,f). At quarter-filling, any state with only horizontal or only vertical dimers clearly cannot be changed by the application of the $\left| A \rangle
\langle \overline{A} \right|$ terms since such terms necessarily involve both horizontal and vertical dimers. Moreover, it is easy to show that \emph{any} ring exchange consisting of alternating occupied and empty bonds will involve an even number of fermions in this state and hence is not allowed. There are $2\times 2^L$ such states. A similar logic dictates that all FPL states like that shown in  Figure~\ref{fig:generic-half} are frozen under any fermionic dynamics: whenever the loops can be sorted into two sets (shown in red and blue) in such a way that an empty bond always connects loops belonging to different sets, all ring exchanges are necessarily even. The number of such states has been counted by us in Ref.~[\onlinecite{Pollmann06c}] and shown to scale as $4^{L+1}$ for large $L$. A number of additional fluctuationless states with the same ``bipartite'' property of loops -- see e.g. Figures~\ref{fig:frozen}(g,h) -- can be constructed, yet we were unable to systematically count them.
Hence it remains an open question whether our RK point is
characterized by a truly extensive GS entropy like that found in
Ref.~[\onlinecite{Misguich03}] or Refs.~[\onlinecite{Fendley05b,Eerten05,Huijse08a,Huijse08b}].

In particular, one wonders whether our model can be actually related to this latter class of supersymmetric models\cite{Fendley05b,Eerten05,Huijse08a,Huijse08b} in any simple way. In short, the supersymmetric Hamiltonians are constructed as
\begin{equation}
H_{\text{SUSY}}=\left(Q+Q^\dag\right)^2 = \left\{Q,Q^\dag\right\}
\label{eq:SUSY}
\end{equation}
where $Q^2={Q^\dag}^2=0$. It has been shown\cite{Fendley05b,Eerten05,Huijse08a,Huijse08b} that if one chooses $Q=\sum_i c_i P_{\langle i \rangle}$, where $P_{\langle i \rangle}$ projects out the states with fermions occupying sites adjacent to $i$, then such Hamiltonians \emph{may} have an exponentially large number of zero-energy states (which are necessarily ground states). This conclusion, however, is sensitive to both the filling fractions and the underlying lattices. While the checkerboard lattice has actually been studied in this context\cite{Eerten05} -- under the name of ``square dimer lattice'' -- our Hamiltonian is clearly different. The ring exchange terms can in fact be produced within this framework by choosing $Q=\sum_{\ll i,j,k,l,m,n\gg} \left(c_i c_k c_m P_j P_l P_n \pm c_j c_l c_n P_i P_k P_m\right)$ where $P_i=1-n_i$; the sum is performed over all ``flippable'' rings consisting of sites $i,j,k,l,m,n$. The resulting supersymmetric Hamiltonian is both local and contains ring exchanges, yet it also contains other terms (including additional one- and two-particle hopping terms). What is the relation between the states of such a Hamiltonian and our Hamiltonian at its RK point is an interesting open question. A related question is whether our Hamiltonian can actually be written in a supersymmetric form by a more judicious choice of $Q$.

\begin{figure}[thb]
  \begin{center}
\subfigure[]{\label{fig:R-state}
\includegraphics[width=0.35\columnwidth]{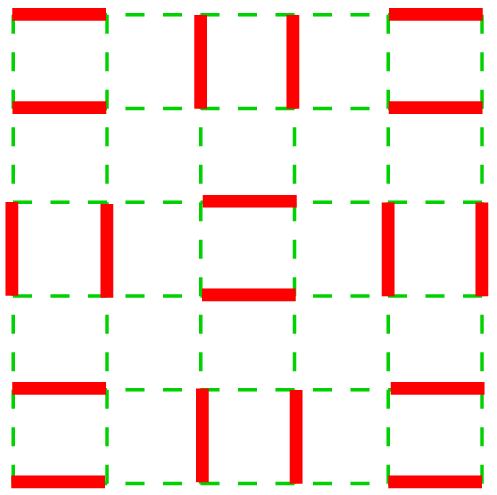}}
$\;\;\;\;\;\;$
\subfigure[]{\label{fig:squiggle}
\includegraphics[width=0.35\columnwidth]{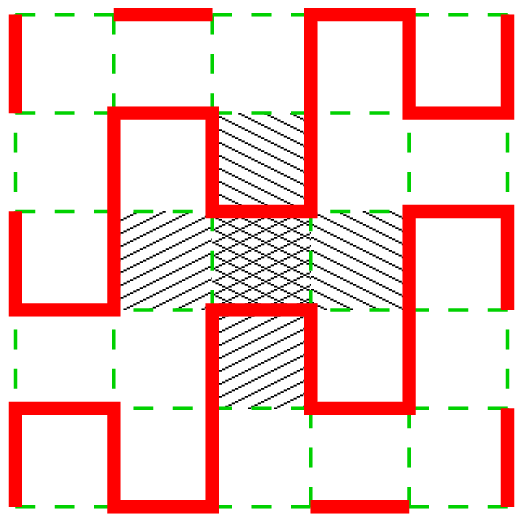}}
\end{center}
\caption{
Fragments of states stabilized by quantum fluctuations:\\
(a) The state maximizing the number of three-fermion ring exchanges at quarter-filling; (b) The ``squiggle'' state that maximizes the number of three-fermion ring exchanges at half-filling. The flippable rectangles involving the central plaquette are shaded.
}
\label{fig:fluctuating}
\end{figure}

Whether or not this is the case, it remains
quite suggestive that the explanation for the large number of additional frozen ground states lies in a more restrictive nature of the fermionic dynamics. We reiterate that these ``new'' frozen states are not related to
any tilt constraints in the height language. The fact the frozen states now include the maximally flat states
(Figure~\ref{fig:frozen}(c,d))  is
particularly interesting since they are supposed to dominate the ordered phase of both quarter-filled and
the half-filled bosonic model for $\mu < \mu_{\text{c}}$ by maximizing the
number of the flippable plaquettes. At quarter-filling, this is a columnar phase quantum dimers; $\mu_{\text{c}} \approx 0.6 g$ numerically.\cite{Syljuasen06b}
At half-filling, this is a conventional
anti-ferroelectric phase of the six-vertex model referred to as the N\'{e}el
phase in Ref.~[\onlinecite{Shannon04}] and as DDW phase in
Refs.~[\onlinecite{Chakravarty02,Syljuasen06a}], its upper boundary has been found to be
$\mu_{\text{c}} \approx -0.374 g$.\cite{Shannon04}
The aforementioned discussion makes it
doubtful that such phases exist in the fermionic model.  At half-filling, the phase dominated by a maximally-flat configuration shown
in Figure~\ref{fig:column-half} appears to be replaced by the somewhat
analogous ``squiggle''  phase shown in Figure~\ref{fig:squiggle}.\cite{Pollmann06b} In the quarter-filled case, the columnar phase appears to be replaced by a two dimensional analogue of the ``R-state'' \cite{Bergman2006,Sikora09} shown in Figure~\ref{fig:R-state}.

\subsection{Excitations at the RK point}
The quantum dynamics of the Hamiltonian $H_{\text{eff}}$ and similarly
$H_{\text{eff}}^{\prime}$ can
be described in terms of a height model as described above. \cite{Henley97} The
associated conserved quantities $\kappa_x$ and $\kappa_y$, as well as their
insensitivity to a constant shift of the height field imply gapless hydrodynamic modes with
$\omega(k)\sim k^2$ as $k\rightarrow 0$. The liquid state of the FPL model
at the RK point corresponds to the rough phase
of the height model, in which case the modes can be identified as the
so-called resonons at wave vector $(\pi, \pi)$ \cite{Rokhsar88} and the
equivalent of the pi0ns, here at $\mathbf  Q = (0,0)$.\cite{Moessner03b,
Moessner03a,Moessner04a} In the following we will elaborate on this physics.

For constructing gapless excitations in our model, we adopt the single-mode
approximation \cite{Feynman72} which has been successfully used for the QDM on the square
lattice.\cite{Rokhsar88,Moessner03b} A trial state with a momentum which differs
from that of the ground state by q is constructed. The momentum is a good quantum
number in a translationally invariant system. Then the variational principle employed
on the states at that momentum yields the basic result that the energy
of the trial state provides an upper bound on the excitation energy at that
momentum. Therefore, as a matter of principle, one can use the SMA to prove gaplessness.
In order to demonstrate the presence of a gap, a different method is
needed.

Let $|0\rangle$ be one of the aforementioned
``liquid'' GS at the RK point. The operator
$d_{\hat{\tau}}^{+(-)}(\mathbf r)$  creates (annihilates) a dimer at position
$\mathbf r$ in direction (polarisation)  $\hat{\tau}$. The density operator
$n_{\hat{\tau}}(\mathbf r)=d_{\hat{\tau}}^+(\mathbf r)d_{\hat{\tau}}^-(\mathbf
r)$ has the Fourier transform $n_{\hat{\tau}}(\mathbf q)=\sum_{\mathbf r}
n_{\hat{\tau}}(\mathbf r) \exp(i\mathbf q\cdot \mathbf r)$. We use the
operators $n_{\hat{\tau}}(\mathbf q)$  to construct the states $|\mathbf q,
\hat{\tau}\rangle=n_{\hat{\tau}}(\mathbf q)|0\rangle$ which are for $\mathbf
q\ne 0$ orthogonal to $|0\rangle$. By using the variational method, the excitation energies $E(\mathbf q, \hat{\tau})$ are bounded by
\begin{equation}
E(\mathbf q, \hat{\tau}) \leq f(\mathbf{q})/S_{\tau \tau}(\mathbf q),
\end{equation}
where
\begin{equation}
f(\mathbf q)=\langle 0|[n_{\hat{\tau}}(-\mathbf q),[H_{\text{eff}}',
n_{\hat{\tau}}(\mathbf q)]]|0\rangle
\end{equation}
\noindent is the oscillator strength and
\begin{equation}
S_{\hat{\tau}\hat{\tau}}(\mathbf q)=\langle 0|n_{\hat{\tau}}(-\mathbf q)
n_{\hat{\tau}}(\mathbf q)|0 \rangle
\end{equation}
\noindent  is the structure factor, essentially the Fourier transformation of the dimer density correlation
function.

It is important that both the oscillator strength and the structure factor can be evaluated as expectation values with respect to the ground state.
The ground state correlations encode information on the excitation spectrum and some of them can be captured by the variational method
which leads to the above expression.
SMA is useful then because there are situations in which gaplessness is present in a generic manner.
The behavior of $f(\mathbf q)$ near $q \rightarrow 0$ can then be used to determine
a bound on the dispersion of the soft excitations. A finite $f(\mathbf q)$ accompanied
by a divergence of the structure factor can be used to infer gapless excitations.
Such a soft mode is a classic signature of incipient order.

In order to
calculate $f(\mathbf q)$, we observe that the density operators commute with
the potential energy term, hence only the kinetic energy term contributes. Let us first consider the case where $\hat{\tau}=\hat x$.
By using  the commutation relation $[d_{\hat{\tau}}^{\pm},n_{\hat{\tau}}]= \mp
d_{\hat{\tau}}^{\pm}$ repeatedly, we compute $f(\mathbf q)$ due to the
resonating terms.
The result for the necessary commutators in order to compute the oscillator strength are (for a square lattice
with lattice spacing $a=1$) reads:
\begin{multline}
[n_{\hat{x}}(-(\mathbf Q + \mathbf
k)),[-gT^{A/B}_{\smallrecv},n_{\hat{x}}((\mathbf Q + \mathbf k))]]\\
= 4g
T^{A/B}_{\smallrecv}
\exp(i \mathbf Q\cdot \mathbf R) \sin^2(\mathbf (\mathbf Q +\mathbf k)
\cdot \hat{y}).
\end{multline}
and
\begin{multline}
[n_{\hat{x}}(-(\mathbf Q + \mathbf
k)),[-gT^{A/B}_{\smallrech},n_{\hat{x}}((\mathbf Q + \mathbf k))]]
\\
= - 4g
T^{A/B}_{\smallrech}
 \exp(i \mathbf Q \cdot \mathbf R) (1 + \cos((\mathbf Q +
\mathbf k) \cdot \hat{y}))
\\
\times (-1 + \cos((\mathbf Q + \mathbf k) \cdot \hat{x}))
\end{multline}

\noindent where $T^{A/B}_{\smallrecv}$ and $T^{A/B}_{\smallrech}$ denote the resonant motion in the minimal ring exchange process with double plaquettes
oriented in a column or  in a  row and with the middle dimer occupied (A kind) or unoccupied (B kind).

If we write  $\mathbf{k}=\mathbf{q}-\mathbf{Q}$ and we consider $\mathbf{q}$ around three different values of $\mathbf{Q} =
(0,0), (\pi,\pi), (0,\pi)$ and therefore small $k$, then we can expand the results for the oscillator strengths.
Then, to leading order we find that $f(\mathbf k)\sim ({\mathbf k} \times
\hat{\tau})^2$ (similar type of expression as the one for the RK model
\cite{Rokhsar88}). Nevertheless, this result alone is not sufficient to give the whole information
of the gapless excitations as the second necessary ingredient, the structure factor, can be also zero
along the direction where $\mathbf{k}$ is zero.

Therefore, let us now consider the structure factor $S_{\hat{x}\hat{x}}(\mathbf q)$ both for the FPL and dimer models independently.
We use the expression first given in Ref.~\onlinecite{Moessner04a}.
Interestingly, the main result is that $S_{\hat{x}\hat{x}}(\mathbf q)\ne 0$ except along the direction $\mathbf
q=(q_x,\pi)$ where it vanishes with the exception of $(\pi,\pi)$. At that point we have the gapless
excitations termed as resonons by RK. At $(0,\pi)$ both $f(\mathbf q)$ and $S_{\hat{x}\hat{x}}(\mathbf q)$ are zero, but their ratio
remains finite. In addition, for the FPL $S_{\hat{x}\hat{x}}(\mathbf q)$ shows no singularities. In that respect, the
FPL model differs from the hardcore dimer model (describing our model at
quarter-filling) for which $S_{\hat{x}\hat{x}}(\mathbf q)$ diverges logarithmically at
$\mathbf Q=(\pi,0)$.\cite{Moessner03b} The difference is due to a slower
algebraic decrease with distance of the correlation function for hardcore
dimer covering.  The correlation function for the dimer model at distance $r$ reads:
\begin{equation}
C_{\hat{x} \hat{x}} \sim  (-1)^{x+y} \frac{y^2-x^2}{r^4} + (-1)^x \frac{1}{r^2}
\end{equation}
\noindent while, for the FPL model it decays with a characteristic exponent which is  higher than 2 \cite{Raghavan97}.

\begin{figure}
(a)\includegraphics[width=80mm]{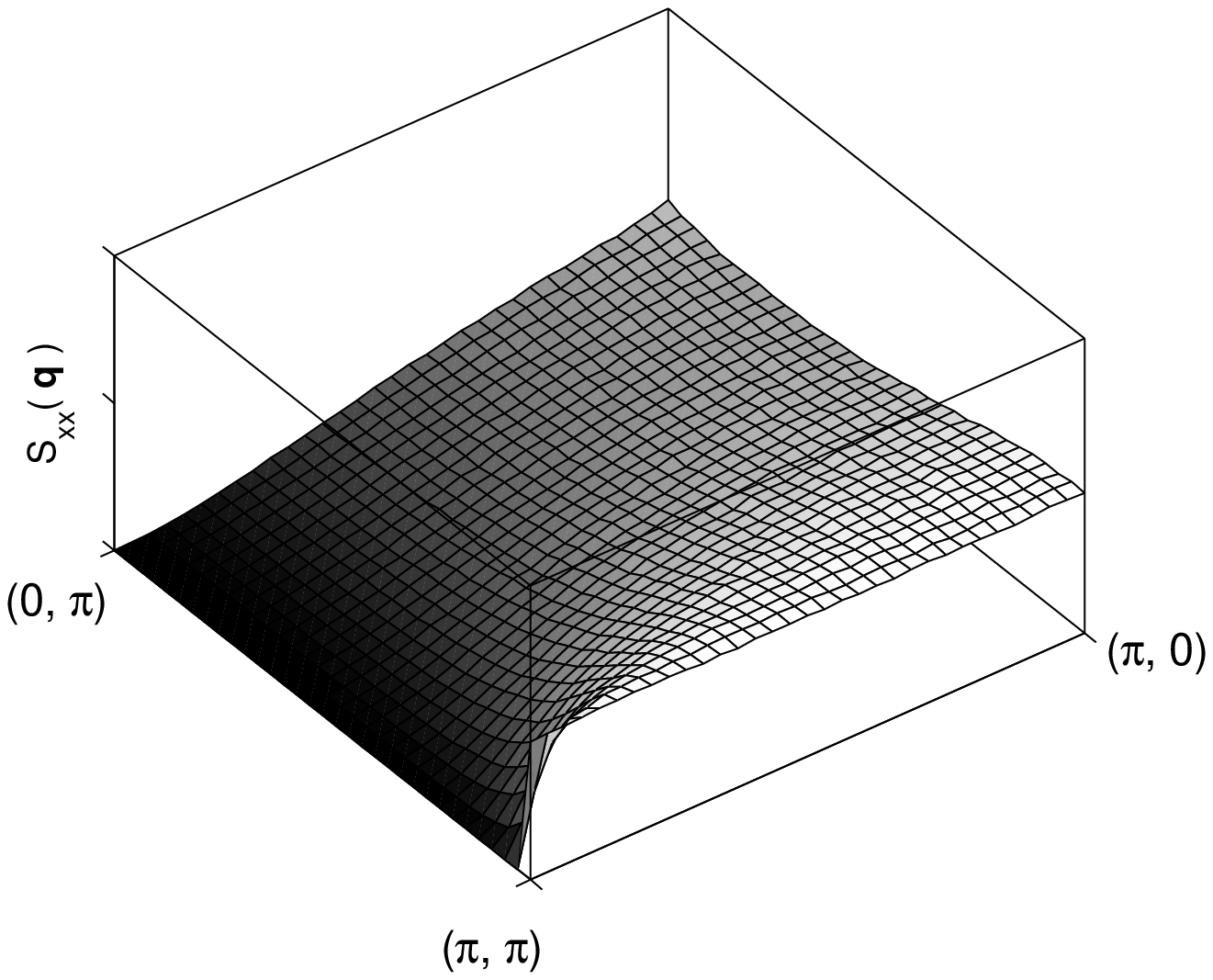}
(b)\includegraphics[width=80mm]{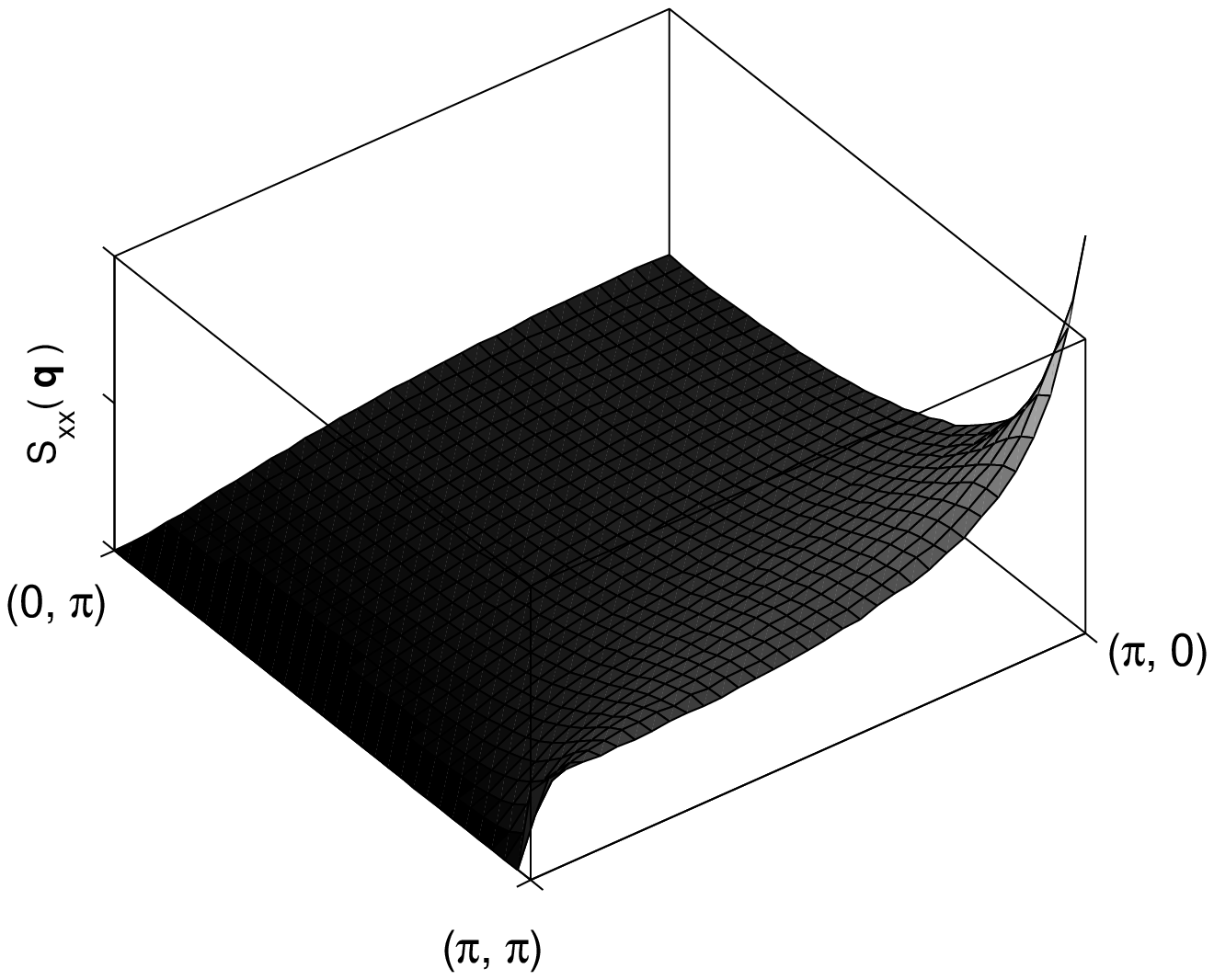}
\caption{Structure factor $S_{\hat{x}\hat{x}}$ for the (a) FPL and (b) hard-core dimers on the square lattice.\label{structurefactor}}
\end{figure}

We have also verified the results for the structure
factors of our model $S_{\hat{x}\hat{x}}(\mathbf q)$ by means of Monte Carlo simulations for both cases, see Figure~\ref{structurefactor}(a) for the
FPL model and Figure~\ref{structurefactor}(b) for the case of hard-core dimers.
The results remain identical (and symmetrical in the indices involved) for the orientation $\hat{\tau}=\hat y$.

\section{U(1) gauge theory}
\label{sec:gauge}

 The effective Hamiltonian (\ref{eq:ringexone}) conserves the number of particles on each crisscrossed square, i.e., the number of loops or dimers touching each site is conserved. This conservation generates a local $U(1)$ gauge invariance, as it is usually the case in models with local constraints [\onlinecite{Fradkin90}]. The gauge structure suggests that we can gain further insight with respect to low-energy excitations by writing our model as a $U(1)$ lattice gauge theory. The usefulness of this approach has already been shown for the quantum dimer model (QDM) on the square lattice \cite{Fradkin90}, strongly correlated proton systems \cite{neto2006} as well as for three dimensional spin systems.\cite{hermele04a} The gauge theory can be derived for the dimer model as well as for the loop model in a very similar manner.
\subsection{Dimer Model}

\begin{figure}
\begin{center}\begin{tabular}{cc}
\includegraphics[height=28mm]{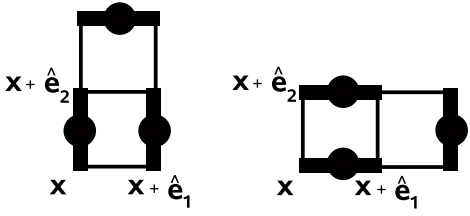}
\end{tabular}\end{center}
\caption{Labeling of links on flipable double-plaquettes at position $\mathbf x$ with unit vectors $\hat{e}_1$ and $\hat{e}_2$.
\label{labeling}}
\end{figure}

We start by defining an integer variable $n_{j}(\mathbf{x})$ for each link $(\mathbf{x},\mathbf{x}+\hat{e}_{j})$ of the square lattice. The  coordinates of a lattice site are denoted by  $\mathbf{x}$, and $\hat{e}_{j=1,2}$ are unit vectors along the axes as shown in Figure~\ref{labeling}.  The states $|\{n_{j}(\mathbf{x})\}\rangle$ span an enlarged Hilbert space which has integer numbers for the links instead only zero or one. We can consider the states $|\{ n_{j}(\mathbf{x})\}\rangle$ as eigenstates of quantum rotor operators $\hat{n}_{j}(\mathbf{x})$ with eigenvalues $n_{j}(\mathbf{x})$.

In order to express the effective Hamiltonian in terms of the $\hat{n}_j(\mathbf x)$, we introduce phases $\hat{\phi}_j(\mathbf{x})\in [0,\pi)$  on the links which are canonical conjugate to $\hat{n}_{j}(\mathbf{x})$. Using the fact that $\exp\left[ \pm i \hat{\phi}_j(\mathbf{x})\right ]$ act as ladder operators, we can write
\begin{eqnarray}
\mathcal{H}_{\mbox{eff}}= U {\sum_{\mathbf{x}, j}}\left (  \hat {n}_j(\mathbf x)-\frac{1}{2}\right) ^2 -2g\sum_{\{\smallrech,\smallrecv\}}\cos \left[\sum \pm \hat \phi \right]. \label{Hgauge}
\end{eqnarray}

Here, the argument of the cosine term contains the sum over phases $\hat {\phi}_j(\mathbf{x})$ with alternating signs around the polygons of perimeter six (double plaquettes). In the limit $U/g\rightarrow \infty$, all ``non-physical'' states are projected out and Eq.~(\ref{Hgauge})  is a faithful representation of the effective Hamiltonian (\ref{eq:ringexone-quarter}).
Next, we introduce staggered gauge and electric fields on the bipartite square lattice $\mathbf{x}=(x_{1},x_{2})$ by
\begin{eqnarray*}
\hat{A}_{j}(\mathbf{x}) & = & (-1)^{x_{1}+x_{2}}\hat{\phi}_{j}(\mathbf{x})\\
\hat{E}_{j}(\mathbf{x})& = &(-1)^{x_{1}+x_{2}}\left(\hat{n}_{j}(\mathbf{x})-\frac12\right),\\
\end{eqnarray*}
The local constraint that each site is touched by exactly one dimer reads
\begin{eqnarray}
\left( \Delta_{j}\hat{E}_{j} \left( \mathbf{x} \right) - \rho(\mathbf{x}) \right) |\mbox{Phys.}\rangle = 0
\label{gauss}
\end{eqnarray}
with the staggered background charge
\begin{eqnarray}
\rho(\mathbf{x})=(-1)^{(x_1+x_2+1)}.
\end{eqnarray}
The lattice divergence is defined as
\begin{eqnarray*}
\Delta_{j}\hat{E}_{j}\left(\mathbf{x}\right) \equiv \hat{E}_{1}(\mathbf{x})-\hat{E}_{1}(\mathbf{x}-\mathbf{e}_{1})+\hat{E}_{2}(\mathbf{x})-\hat{E}_{2}(\mathbf{x}-\mathbf{e}_{2}). \end{eqnarray*}
The dimer hardcore dimer constraint is now reflected by the standard Gauss' law (\ref{gauss}) in the presence of a staggered background charge density. Using the staggered fields as defined above, the Hamiltonian (\ref{Hgauge}) in reads
\begin{eqnarray}
\mathcal{H}_{\mbox{eff}}=U{\sum_{\mathbf{x}, j}}\hat{E}_{j}^2(\mathbf{x})-2g\sum_{\mathbf{x}} \cos \left[\sum_{\orients} \hat{A}_{j}\left(\mathbf{x}\right) \right].\label{Hqed}
\end{eqnarray}
The argument of the cosine term denotes the oriented sum of staggered vector potentials $\hat{A}_{j}\left(\mathbf{x}\right)$  around double plaquettes.

\subsection{Loop model}

Let us now consider the loop model which corresponds to the half filled  checkerboard lattice. The matrix-elements of the effective Hamiltonian (\ref{eq:ringexone}) have different signs depending on whether the link in the middle is occupied or not. Using the gauge transformation presented in section \ref{sign}, we can rewrite the Hamiltonian in a representation which has only negative matrix elements and redo the derivation presented above for the dimer model. The only difference between the dimer and loop model is the form of Gauss' law which reflects the loop constraint. Using the same notation as for the dimer model, the local constraint that each site is touched by exactly \emph{two} dimer reads
\begin{eqnarray}
\Delta_{j}\hat{E}_{j} |\mbox{Phys.}\rangle = 0,
\label{gauss2}
\end{eqnarray}
Thus the corresponding Gauss' law has no background charges.\\

Eq.  (\ref{Hqed}) has similarities with the Hamiltonian of the compact quantum electrodynamic (QED) in $2+1$ dimensions in which the considered charges correspond to fractional charges of $e/2$.\cite{polyakov77} Polyakov showed for the compact QED in $2+1$ dimension that it has an unique and gapped ground state. Two charges are confined and the energy grows linearly with the distance between the two charges. Our model shows important differences. The fields $E_{j}(\mathbf{x})$ are half integers instead of integers and the constraint selects configurations with a background charge $\rho(\mathbf{x})$. This leads to a frustration which is reflected by the macroscopic degeneracy of the classical ground states. Furthermore, the definition the flux in the cosine term differs.

The formulation (\ref{Hqed}) provides an excellent starting point for further systematic investigations. The plaquette duality transformation allows to map the Hamiltonian to a height model and to use path integrals for a detailed study of the ground-state as well as low-energy excitations \cite{Fradkin90}.

\section{Conclusion}
\label{sec:conclusion}

In this study, we analyzed in detail a model of strongly correlated spinless fermions on a checkerboard lattice both at half and quarter-filling by mapping it onto a quantum dimer and a quantum fully-packed loop models on the square lattice respectively. The lowest-order effective Hamiltonian  is given by ring hopping processed of three fermions around hexagons. The symmetries and conservations of the model were studied in detail. We found a large number of frozen states which are not related to any height constraints and specific for the fermionic case. Using a topological and an algebraic approach, we showed that the sign problem, induced by the fermion statistics in the half filled case, can be cured. We explained and studied different aspects of this observation and showed that it does not remain true when further ring exchange processes are also taken into account. By adding an extra term to the Hamiltonian, we fine tuned the system to the Rokhsar-Kivelson point where the ground states are known exactly. Using the proposed sign gauging and the single mode approximation, we argue that the Rokhsar-Kivelson point is a quantum critical point and the system crystalizes once it is slightly detuned. Moreover we mapped the system at both fillings to the corresponding gauge theories which dictates confinement of fractionalized defects created by an addition (or subtraction) of fermions.

Recent studies of other lattices such as kagome at 1/3 fermionic filling concluded that the defects (in the form of fractional charges)
are confined.\cite{Obrien2010} This is also the case in the present study and it is a consequence of the confined nature of the gauge theory as explained above.
The fact that the lattice under consideration is bipartite as well as the imposed local constraints, lead to the fractionalization of the defects with protected value of the fractional charge (one half in we dope the system with an extra fermion).

Finally, we should caution the reader who may be left with an impression that aside from the fact that only odd ring exchanges are present, the fermionic nature of the model considered here can be essentially ``gauged away''. Therefore it must be reiterated that the methods we have proposed are of somewhat limited applicability. Firstly, they only work for special filling fractions. Their non-generic nature is obvious from the fact that for the quarter-filled case, a simple local gauge transformation would suffice to make the quantum dynamics of the Frobenius type, while for the half-filled case a far more complicated non-local transformation was required. Therefore, it would be extremely interesting to study this model away from these special fillings and try to identify the features that are generic.

\begin{acknowledgments}
The authors would like to thank P.~Fendley, R.~Moessner, D.~Poilblanc and E.~Runge for many
illuminating discussions.
F.~P.\ has benefited from the hospitality of IDRIS, Orsay during a stay made
possible by a HPC Europe grant (RII3-CT-2003-506079). K.~S.\ is supported in part by the NSF under grant DMR-0748925. J.~B.\ and K.~S.\ are grateful for
the hospitality of the MPIPKS, Dresden and the Instituut-Lorentz, Leiden.
\end{acknowledgments}


\end{document}